# Evidence for Possible Clouds in Pluto's Present Day Atmosphere
# Running Head: Clouds in Pluto's Atmosphere

S.A. Stern, J.A. Kammer, E.L. Barth, K.N. Singer, T.R. Lauer, J.D. Hofgartner, H.A. Weaver, K. Ennico, C.B. Olkin, L.A. Young, the New Horizons LORRI Instrument Team, the New Horizons Ralph Instrument Team, and the New Horizons Atmospheres Investigation Team

Author Addresses (and email for lead author):
Stern, S.A., Southwest Research Institute, Suite 300, 1050 Walnut St, Boulder, CO 80302, astern@swri.edu
Barth, E.L., Southwest Research Institute, Suite 300, 1050 Walnut St, Boulder, CO 80302
Singer, K.N., Southwest Research Institute, Suite 300, 1050 Walnut St, Boulder, CO 80302
Lauer, T.R., The National Optical Astronomy Observatory, P.O. Box 26732, Tucson, AZ 85726
Hofgartner, J.D., Jet Propulsion Laboratory, 4800 Oak Grove Dr. MS 183-40, Pasadena, CA 91109
Weaver, H.A., Johns Hopkins University Applied Physics Laboratory, Space Exploration Sector 11100 Johns Hopkins Road, Laurel, MD 20723
Ennico, K., National Aeronautics and Space Administration (NASA) Ames Research Center, Space Science Division, Mail Stop 245-6, Moffett Field, CA 94035
Olkin, C.B. Southwest Research Institute, Suite 300, 1050 Walnut St, Boulder, CO 80302
L.A. Young, Southwest Research Institute, Suite 300, 1050 Walnut St, Boulder, CO 80302

**Abstract**

Several trace constituents can reach saturation vapor pressure in Pluto's present day atmosphere. As such, we describe a search for discrete cloud features in Pluto's atmosphere using New Horizons data obtained on 14-15 July 2015, during the Pluto flyby closest approach. We report that Pluto's present day atmosphere is at least largely (>99% by surface area) free of discrete clouds. We also report a handful of features that may plausibly be clouds, all of which were detected obliquely and at high phase angle observing geometry. No cloud candidates were identified away from the terminators or in low phase (backscattering geometry) images.

**1. Introduction**

Pluto's atmosphere has long been known to contain a variety of minor constituents (e.g., $CH_4$, CO) beyond the $N_2$ that dominates its composition (e.g., Stern et al. 2015). New Horizons has recently identified trace constituents including $C_2H_2$, $C_2H_4$, and $C_2H_6$, as well as extensive haze layers (Stern et al. 2015; Gladstone et al. 2016; Young et al. 2017). Submillimeter observations by ALMA have also recently reported HCN as another atmospheric trace constituent (Lellouch et al. 2017).

Numerous planets in our solar system, including Venus, Earth, Mars, Titan, and all four of the giant planets, possess atmospheres that contain clouds, i.e., discrete atmospheric condensation structures. This said, it has long been known that Pluto's current atmosphere is not extensively cloudy at optical or infrared wavelengths. Evidence for this came primarily from the high amplitude and temporal stability of Pluto's lightcurve (Stern 1992). However, because no high spatial resolution imagery of Pluto was possible before New Horizons, it remained to be seen if clouds occur over a small fraction of Pluto's surface area.

We, therefore, undertook cloud searches against the disk, as well as near and on Pluto's limb, using imagery obtained during the Pluto flyby. Here we report the results of these searches, as well as new analysis of currently known species that can condense at various altitudes in Pluto's Present Day Atmosphere (PDA) using temperature and pressure profiles retrieved by New Horizons (Gladstone et al. 2016; Hinson et al. 2017; Young et al. 2017). As we discuss below, our searches included the entire close encounter hemisphere of Pluto. Both the limb and the disk of Pluto itself were searched for cloud like features. Seven bright, discrete, possible cloud candidates were identified, all at oblique geometries near the terminators. Unfortunately, no stereo information is available to demonstrate that any of the identified candidates are in fact atmosphere features as opposed to surface features.

This paper begins in §2 with a discussion of the condensation conditions needed to create clouds in Pluto's PDA; §3 then describes our cloud searches and their findings; §4 discusses the common attributes of cloud candidates and points to future work to advance knowledge on this front; §5 summarizes our conclusions.

## 2. Pluto Cloud Condensation Modeling

Atmospheric dynamics, radiation, and latent heat all play a role in the formation of clouds. The degree to which one process dominates over the others determines the nature of the clouds that are formed, e.g., fog vs. cumulus vs. stratus, etc.

Physical mechanisms that can trigger cloud formation derive from surface heating, flow over mountains and other topography, pressure systems, and weather fronts. Strobel et al. (1996) estimated the radiative timescale near Pluto's surface to be ~10-15 terrestrial years. This long timescale would seem to indicate that radiation plays a small or negligible role in triggering cloud formation near Pluto's surface. Our current knowledge of winds near Pluto's surface derives primarily from general circulation models (GCMs; e.g., Forget et al. 2017; Toigo et al. 2015), which predict winds of a few m s$^{-1}$ near the surface induced by both flow over topography and the $N_2$ surface condensation/sublimation cycle. Winds of a few m s$^{-1}$ result in dynamical timescales ~10 terrestrial days. So on Pluto today, flow over topography and the pressure/wind changes associated with the nitrogen sublimation cycle are expected to be key triggers for cloud formation.

In order to truly understand if clouds are a limited or widespread phenomenon on Pluto, we would need longer-term temporal coverage, such as a future orbiter mission could provide. In the current absence of a dedicated orbiter, GCMs and other models will continue to be important to make predictions.

With the knowledge we have gained from the New Horizons mission, we can examine the possibility of cloud formation in Pluto's lower atmosphere by considering, (1) if any of its atmospheric gases are abundant enough to condense, and (2) whether cloud particle growth will proceed on a reasonable timescale.

To address the first issue, the measured abundance of each known species detected in Pluto's atmosphere can be used to calculate a condensation temperature – the temperature at which the species reaches saturation (i.e., where the partial pressure equals the saturation vapor pressure). Comparing these species condensation temperatures to Pluto's atmospheric vertical temperature profiles gives an estimate the altitude ranges (if any) that a given species can reach saturation. Figure 1 shows such condensation temperature curves overlaid on the New Horizons entry and exit vertical atmospheric temperature profiles (Hinson et al. 2017).

The abundances used to construct the saturation profiles in Figure 1 were taken from Gladstone et al. (2016) and Young et al. (2017) for $CH_4$, $C_2H_4$, $C_2H_6$, $C_2H_2$, and Lellouch et al. (2017) for CO and HCN. As Figure 1 reveals, only $C_2H_6$, $C_2H_2$, and HCN reach saturation for both temperature profiles. In contrast, $CH_4$ and $C_2H_4$ require the colder temperatures at the location of the entry profile to reach saturation. In all cases, once saturation is reached the vapor remains in a supersaturated state down to Pluto's surface. Note that although CO is more volatile than the various hydrocarbon species and HCN, the abundance of CO is too low for it to condense in Pluto's atmosphere.

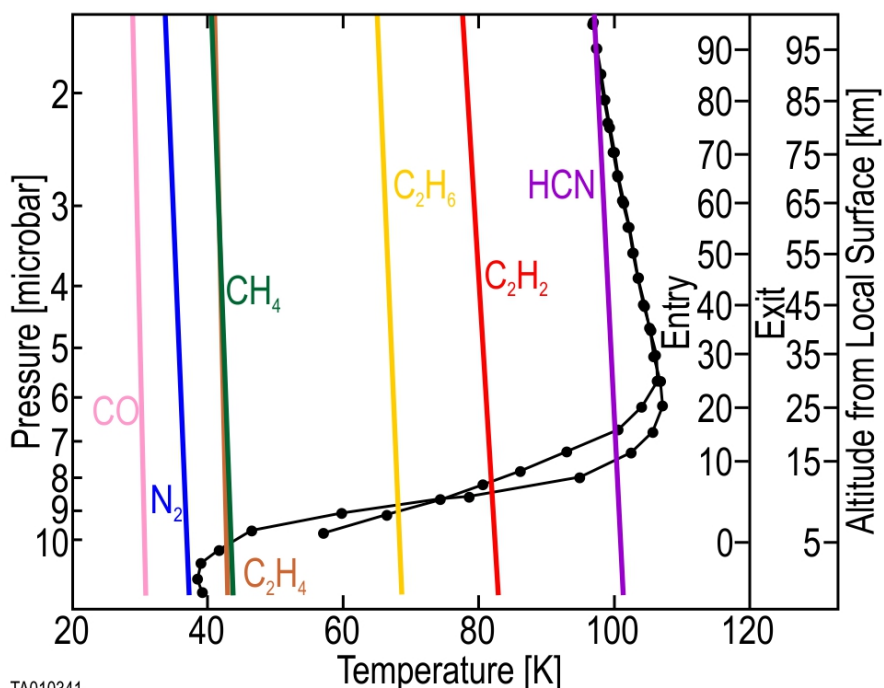

Figure 1. Condensation curves for the major and minor species detected in Pluto's atmosphere as a function of altitude and atmospheric pressure. The New Horizons entry and exit temperature profiles are shown as black curves; the entry profile has a higher surface pressure, which lies at a lower topographic altitude (Hinson et al. 2017). Saturation occurs for each species at and below the pressure level where its saturation profile (the approximately vertical colored line for each species) intersects Pluto's atmospheric vertical temperature profile.

We now discuss the second issue noted above, cloud particle growth. Rannou & Durry (2009; hereafter RD09) examined the nucleation of $N_2$, CO, and $CH_4$ in Pluto's atmosphere using pre-New Horizons occultation temperature profiles and estimated abundances. They concluded that $CH_4$ reaches saturation at the lowest altitudes, followed at successively higher altitudes by CO and $N_2$. However their assumed CO abundance was much higher than was later observed by ALMA (Lellouch et al. 2017). Lellouch et al. (2009; hereafter L09) also discussed the supersaturation of $CH_4$ using stellar occultation data. Their estimate of vapor pressures 30% above saturation (i.e., a supersaturation of 0.3) near the surface is consistent with the profiles shown in Figure 1. Neither RD09 nor L09, however, modeled the microphysics of the cloud formation process, which we consider next.

Nucleation is the initial step in the formation of clouds. We calculate nucleation rates using the classical heterogeneous nucleation theory equations of Pruppacher & Klett (1997). Heterogeneous nucleation requires the presence of particles to act as condensation nuclei. We assume that the haze observed by New Horizons supplies these condensation nuclei, which Gladstone et al. (2016) estimated at a number density of ~0.8 $cm^{-3}$ near the surface.

Because haze is seen in Pluto's present day atmosphere, heterogeneous nucleation is a more plausible means to initiate cloud formation than homogeneous nucleation (which implies no preexisting condensation nuclei). Both processes involve overcoming an energy barrier, but the presence of preexisting condensation nuclei reduces the amount of energy needed as a function of the "contact parameter." This contact parameter describes the relationship between the condensing species and the substrate (i.e., the condensation nuclei), and so is a function of the composition of both. The contact parameter is determined from lab measurements that determine the saturation value at which the nucleation rate reaches unity, i.e., the critical saturation for nucleation. Unfortunately, very few such measurements exist for non-terrestrial condensing species but important exceptions for our purposes include methane, ethane, and butane nucleation onto a number of substrates, including Titan tholins (Curtis et al. 2005, 2008).

Barth (2017) described how these measurements were used to estimate a critical (i.e., cloud forming) relative humidity for a number of hydrocarbon and nitrile species in Titan's atmosphere. We follow that same methodology here, using a critical relative humidity of 110% for methane and choosing a mid-range value of 130% for all other species.

Once a stable cloud particle is formed, it can grow in a supersaturated environment through condensation. Because condensation is directly correlated to the atmospheric temperature, growth proceeds more slowly at colder temperatures.

The cold temperatures near Pluto's surface make condensation very inefficient there, with characteristic timescales to grow a 2 μm radius particle of order ~$10^6$-$10^9$ terrestrial years for all condensing species we considered except methane. This means that all nucleation and growth by condensation on any (non-methane) cloud particles in Pluto's atmosphere must be occurring at higher altitudes (e.g., ~8-10 km above the surface at the entry site) where temperatures are >~60 K, and the characteristic condensation timescales are short compared to a Pluto year.

We simulated cloud particle formation and evolution in Pluto's atmosphere using the Pluto module of CARMA, the Community Aerosol and Radiation Model for Atmospheres (Turco et al. 1979; Barth 2017). A Pluto atmosphere was created using the temperature-pressure (T-P) profiles from Hinson et al. (2017); these are shown as the black T-P profiles in Figure 1. CARMA's aerosol model simulates nucleation, condensation, evaporation, sedimentation, and coagulation of atmospheric particles. Particles are represented in discrete size bins in a column of atmosphere partitioned into layers.

As will be seen below in §3, all of the suspected cloud features detected in New Horizons imagery are near Pluto's surface, so we focus our CARMA simulations to altitudes below 25 km. Such simulations were conducted for $CH_4$, $C_2H_2$, $C_2H_4$, $C_2H_6$, and HCN using atmospheric columns representing both the entry and exit sounding sites (entry only for $CH_4$ and $C_2H_4$). Most species were added to the model using the vapor pressure and latent heat references shown in Barth (2017, Table 2); however, the $CH_4$ vapor pressure equations and latent heat are calculated from Moses et al. (1992), the $C_2H_4$ vapor pressures are calculated from Brown & Ziegler (1980) and Lara et al. (1996) for latent heat of sublimation. In all cases, the vertical grid extended from the surface to 25 km altitude with a grid spacing of 500 m. Haze particles were initialized in each altitude layer in the model using a log-normal size distribution (typical for atmospheric aerosols) consisting of 35 size bins that double in mass with each step; the number of particles peaks in size at 0.1 micron radius (also see Gladstone et al. 2016), as shown in Figure 2. The particles have a fractal shape with a monomer radius of 50 nm and a fractal dimension of 2. The haze particles are meant to be representative of the aerosols observed in Pluto's atmosphere but are not themselves microphysically modeled. Variation of the parameters listed above does not alter the conclusions with regard to ice particle formation described below. The haze particles were then subjected to vertical transport (sedimentation and eddy diffusion), including a flux into the top layer to maintain a constant supply of condensation nuclei. All particles in the model can fall out the bottom layer to the surface.

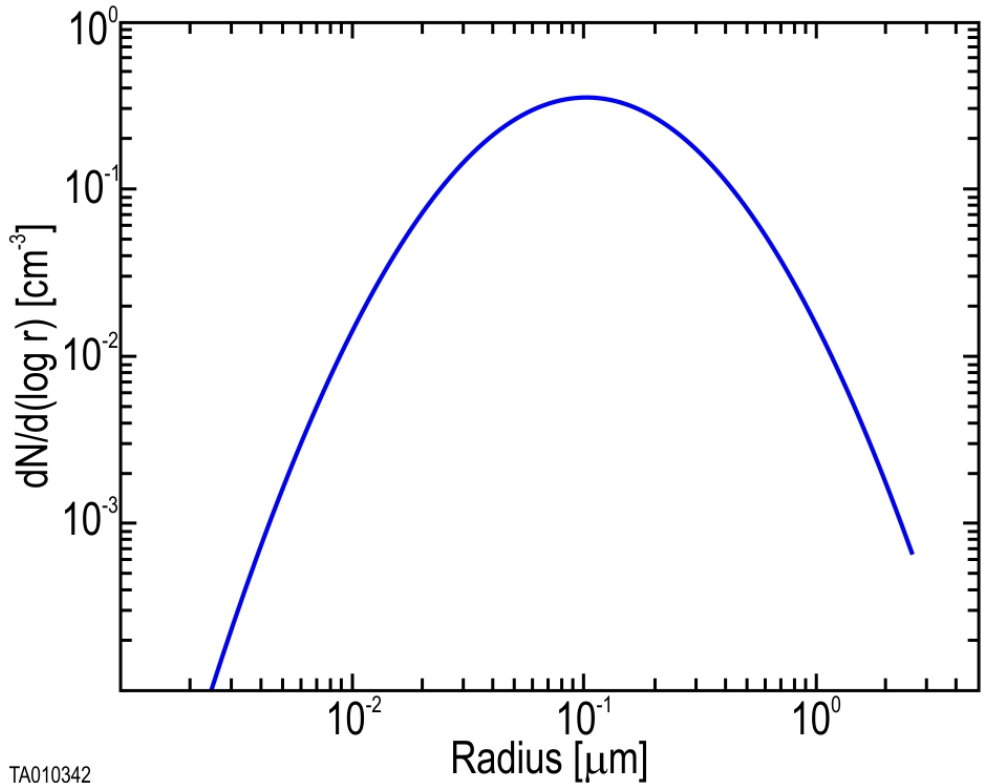
Figure 2. Assumed size distribution of Pluto haze particles that serve as condensation nuclei in our cloud condensation model. This log-normal distribution peaks at a radius of 0.1 micron and sums to 0.8 cm$^{-3}$, consistent with Gladstone et al. (2016).

In all our simulations, cloud formation occurred near and below the condensation altitudes shown in Figure 1. Figure 3 shows the results of our cloud formation modeling calculations, and compares cloud top heights (the knee in each profile) and species abundances for the two temperature cases. For both entry and exit, HCN nucleation occurs at the highest altitudes. The similarity in entry and exit HCN number density profiles is due to the similar slope in the two temperature profiles at these altitudes. Although $C_2H_4$ nucleation does appear to occur as indicated in the entry simulation, the process is inefficient and there are too few cloud particles to contribute to any observed cloud opacity. $CH_4$ and $C_2H_2$ nucleation produces a similar total number of particles; however, as described below, $CH_4$ particles grow to be the larger of the two types. $C_2H_6$ nucleation creates by far the most cloud particles and so would be the dominant constituent in any visible cloud.

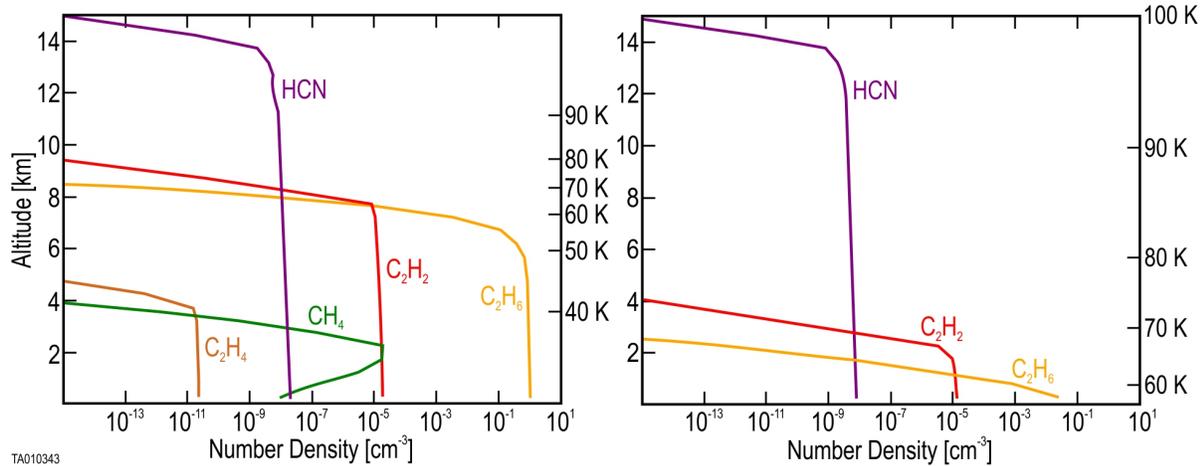

Figure 3. Pluto cloud particle number density profiles from CARMA simulations. The left hand plot corresponds to the entry temperature profile; the right hand plot corresponds to the exit temperature profile. The corresponding temperature levels are indicated on the right hand side of each plot.

More insight into the processes controlling the relative abundances of cloud particles shown in Figure 3 can be gained from looking at the nucleation rates and condensation timescales for each species. To facilitate this comparison, temperature will be used as a proxy for altitude (the mapping between temperature and altitude above Pluto's surface is shown in Figure 3). In the case of $CH_4$, nucleation rates are high even for the cold temperatures near Pluto's ~40 K surface. In contrast, $C_2H_4$ nucleation is inefficient (i.e., slow) for cold temperatures, and so it does not provide a significant source of cloud particles. The vapor pressure of $CH_4$ at 40-50 K is also a factor of ~$10^{10}$ times higher than both $C_2H_6$ and $C_2H_4$, resulting in a much more rapid growth timescale, allowing particles to grow to radii greater than 10 μm (with a peak in their size distribution near 4 μm). The decrease in the $CH_4$ density profile near the surface is due to precipitation of larger $CH_4$ particles, whose fall timescale becomes shorter than their formation timescale. (Fall timescales for ice particles over a distance of 2 km are larger than 1 terrestrial day for 1 μm particles, about 12 hours for 2 μm particles, 6 hours for 4 μm particles, and less than 3 hours for 10 μm particles.) At 60-70 K, the nucleation rate for $C_2H_6$ is greater than that of $CH_4$ at 40 K, and thus overall there are significantly more ethane than methane cloud particles formed (for ethane, saturation is reached on Pluto at an altitude level where both temperature profiles are ~70 K; for $CH_4$ it is reached on Pluto only at 40 K.).

For both $C_2H_6$ and $C_2H_2$, nucleation and further growth through condensation are only efficient at temperatures above ~50-60 K. Because the temperature gradient is steep near the altitude where they begin forming cloud particles, these processes only occur within a narrow altitude range of a few kilometers. For the case of HCN, nucleation is only efficient in the 0-25 km altitude domain modeled because the saturation level is reached at the much higher temperature of 100 K, which only occurs at higher altitudes. Condensation growth of HCN is inefficient, with timescales ~$10^9$ seconds or greater, even with its higher level of supersaturation. The size of the HCN particles is thus controlled by the size of the condensation nuclei present, i.e. HCN will form a cloud particle but will not grow further by condensation.

Another factor that can limit the number of cloud particles formed is the number of haze particles available to serve as condensation nuclei. The haze particle population was checked for each of the above cases to see if the number used (0.8 cm$^{-3}$) limited the number of cloud particles. The only species that showed a noticeable effect on reducing the number of haze particles was $C_2H_6$. However, a subsequent test with an increased number of haze particles only showed modest changes in the $C_2H_6$ cloud particle abundance, so the major limiting factor for the number of cloud particles formed is the effect of cold temperature on the nucleation rate.

These simulations show the altitudes where cloud particles can be expected to be found in Pluto's lower atmosphere, and can be used to estimate the relative abundances of each species. Our column (i.e., 1-D) microphysics model is not designed to explicitly simulate the atmospheric dynamical motions that contribute to optically thick, horizontally extended clouds. However, some conclusions can be drawn through modeling a perturbation of the system. For example, adding a 1 m s$^{-1}$ updraft in the bottom 5 km of the $C_2H_6$ entry profile increased the (geometric limit) optical depth at visible wavelengths from 10$^{-4}$ to ~0.5, and a similar test of $CH_4$ results in τ~0.05 (from 10$^{-6}$ in the unperturbed case). Thus, small perturbations to the observed atmospheric column can promote formation of optically thick clouds.

While vertical motion is less well constrained, there is certainly topography to induce an upslope flow, so the updraft simulations described above are not unreasonable. We also note that at least two of the cloud candidates described below in §3 appear to be in the vicinity of mountainous terrain. Finally, we note that Forget et al. (2017) concluded that $CH_4$ ice clouds could form at altitudes within 1 km of the surface with horizontal extents >~100 km (the only volatiles they consider, CO and $N_2$, do not atmospherically condense in their GCM). Their conclusions on the potential presence of $CH_4$ clouds are based on supersaturation, but combined with our CARMA simulations and calculations of nucleation and growth timescales, we see that localized cloud formation is possible in Pluto's present day atmosphere.

### 3. Pluto Cloud Searches and Findings

**3.1 Selection of cloud search images**. Our search for clouds on Pluto utilized panchromatic visible light images from both imagers onboard New Horizons: the Long Range Reconnaissance Imager (LORRI; Cheng et al 2008), and the Multispectral Visible Imaging Camera (MVIC; Reuter et al. 2008). A total of 41 LORRI and MVIC images were examined for clouds, all include both Pluto surface and sky above the limb.

We visually examined images taken at times ranging from the hours immediately following closest encounter with Pluto to several (terrestrial) days later. This time frame was chosen based on high-phase angle imaging geometry, which enabled searches for forward scattered light from possible clouds in the atmosphere.[1] The images we searched are enumerated in Table 1; the "horizontal resolution" column is measured at the limb, and is the same as the vertical resolution.

Development of an automated process to search for clouds on Pluto is a non-trivial task, particularly without a sample of known cloud features to establish a baseline for detection. The authors therefore chose to examine the image sets by eye, adjusting the brightness and contrast scale as needed depending on the particular exposure time and solar illumination of each case. Additionally, higher resolution LORRI images were processed using the Lucy (1974) – Richardson (1972) algorithm, a point spread function (PSF) deconvolution technique that improves the visual detection of fine-scale features, which has been used and described previously for New Horizons images by Weaver et al. (2016). Visual inspections were performed on all images by at least two and often more authors of this paper. During visual examination, individual authors flagged potential cloud candidates based on a common set of qualitative criteria (see next subsection). Almost all images inspected were found by multiple authors to be devoid of any candidate cloud candidates.

**Table 1: Pluto Images Examined for Cloud Candidates**

| Image ID | UTC Start-End (DOY H:M:S) | Solar Phase Angle (deg) | Horizontal Resolution (km pix$^{-1}$) |
|---|---|---|---|
| PEMV_01_P_MVIC_LORRI_CA | 195 11:31:05 – 195 11:38:08 | 63.5 | 0.31 |
| PEMV_01_P_HiPhase_HiRes | 195 12:00:41 – 195 12:05:53 | 146.3 | 0.33 |
| PEMV_01_P_CharonLight | 195 12:08:13 – 195 12:13:43 | 157.5 | 0.41 |
| PELR_P_DEP_SOONEST | 195 15:04:21 – 195 15:08:49 | 169.7 | 0.80 |
| PELR_P_MULTI_DEP_LONG_1 | 195 15:41:50 – 195 15:46:50 | 169.0 | 0.96 |
| PELR_P_LORRI_DEP_0 | 195 18:26:03 – 195 18:32:31 | 167.3 | 1.6 |
| PELR_PC_MULTI_DEP_LONG_2_01 | 195 19:04:45 – 195 19:07:32 | 167.1 | 1.8 |
| PELR_PC_MULTI_DEP_LONG_2_04 | 195 19:35:00 – 195 19:41:24 | 167.0 | 1.9 |
| PELR_PC_MULTI_DEP_LONG_2_05 | 195 19:44:00 – 195 19:46:40 | 166.9 | 1.9 |
| PELR_P_LORRI_ALICE_DEP_1 | 195 21:02:43 – | 166.6 | 2.3 |

---

[1]However we note that casual inspections of images taken at lower phase angles on approach to Pluto revealed no evidence of cloud candidates.

| | 195 21:09:16 | | |
|---|---|---|---|
| PELR_P_LORRI_ALICE_DEP_2 | 195 22:09:15 – 195 22:13:58 | 166.4 | 2.5 |
| PELR_P_LORRI_ALICE_DEP_3 | 196 00:03:20 – 196 00:08:03 | 166.2 | 3.0 |
| PELR_PC_MULTI_DEP_LONG_3_01 | 196 03:04:00 – 196 03:04:30 | 166.0 | 3.7 |
| PELR_PC_MULTI_DEP_LONG_3_04 | 196 03:24:12 – 196 03:24:27 | 165.9 | 3.8 |
| PELR_P_LORRI_FULLFRAME_DEP | 196 03:26:40 – 196 03:28:10 | 165.9 | 3.8 |
| PELR_P_LORRI_ALICE_DEP_4 | 196 07:02:00 – 196 07:03:30 | 165.8 | 4.7 |

**3.2 Selection of cloud candidates**. Of all the images we examined, we found that only a few contained possible clouds or cloud-like features. These Possible Cloud Candidate (PCC) features were identified as having three qualitative criteria: (i) bright features relative to surrounding terrain, (ii) cloud-like 'fuzziness' or diffuse feature appearance, and (iii) discrete areal extent. Only cloud features detected by these criteria by multiple authors were designated as PCCs. This yielded the seven features shown in Figure 4. We note that several of these PCCs were identified in multiple images. No other cloud-like features that meet all three of the criteria given above were identified by multiple authors.

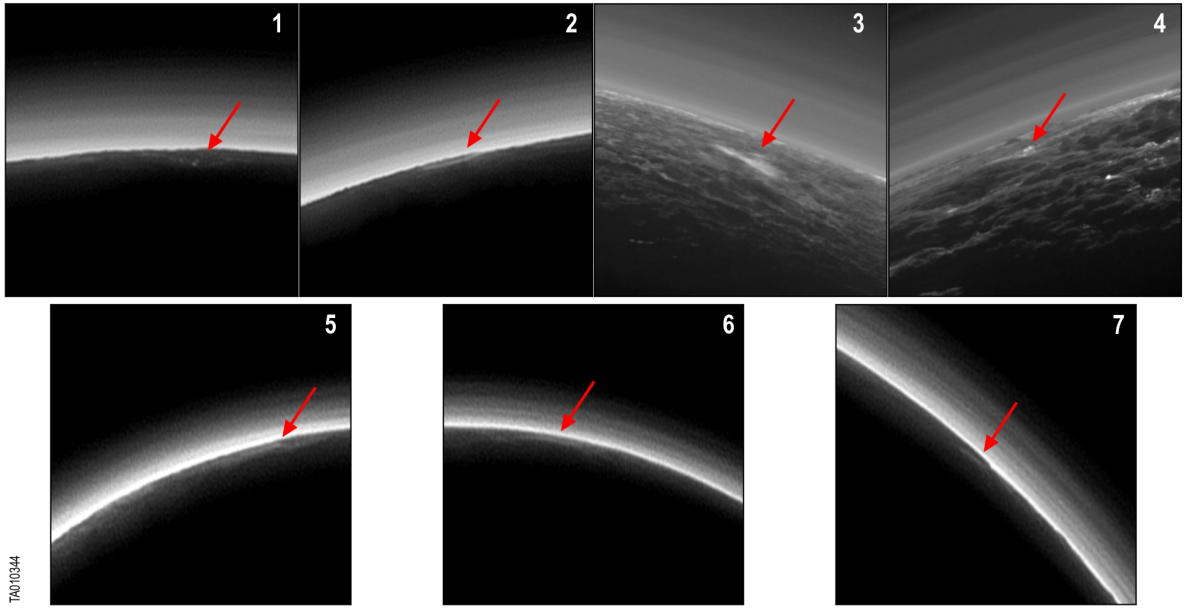

Figure 4. Seven Possible Cloud Candidates (PCCs) identified in this work. Two of these images (3, 4) were taken by MVIC; the other five (1, 2, 5, 6, 7) were taken by LORRI. Arrows indicate each PCC.

**PCC properties, locations, and local times**. Each of our seven PCCs was detected in high-phase imagery, and seen near the limb of Pluto. Most appeared as bright or elongated features (2, 3, 5, 6, 7), but two of the candidates (1, 4) appeared instead as clusters of bright spots. Horizontal extents of the PCCs, i.e., their dimension parallel to Pluto's surface, varied, with candidate 1 being less than 10 km long, while the others ranged from 30 to 70 km, as shown in Table 2. We found that the linear PCCs tended to have similar, narrow aspect ratios, and were generally located in or around ridges or other topography. Because no stereo imagery is available of any of the PCCs, we could not estimate PCC heights or prove that they are located above the surface; however, the fact that all our PCCs appear to be low-lying is in accord with the predictions of cloud condensation models discussed in §2, which suggest formation at altitudes below ~10 km.

Estimated PCC brightness (measured as I/F, where I is the scattered intensity and F is the incident solar flux) is also provided in Table 2. Pluto's global haze layers have similar brightness, i.e., I/F~0.2-0.3 in LORRI images made at similar phase angles of 165º – 169º (Gladstone et al. 2016). Gladstone at al. (2016) used the blue color and high forward scattering behavior of the hazes to estimate their line-of-sight optical depth to be $\tau$~0.16 and their number density to be n~0.8 particles cm$^{-3}$. The fact that we detected the PCCs only in high phase angle images is consistent with their being preferentially forward scattering. However, because the PCCs were only observed in forward scattering we cannot independently verify this. If the PCCs have a similar composition and particle size distribution as the hazes then their similar I/F's may indicate a similar optical depth and number density.

Table 2: Cloud Candidate Properties, Sizes, and Brightness

| PCC | Image ID | General Description | Long Dimension (km) | I/F |
|---|---|---|---|---|
| 1 | PELR_P_MULTI_DEP_LONG_1 | Bright spots | <10 | ~0.2 |
| 2 | PELR_P_MULTI_DEP_LONG_1 | Linear | ~55 | ~0.3 |
| 3 | PEMV_01_P_CharonLight | Linear | ~70 | ~0.1 |
| 4 | PEMV_01_P_CharonLight | Bright spots | ~30 | ~0.1 |
| 5 | PELR_P_LORRI_ALICE_DEP_3 | Linear | ~30 | ~0.1 |
| 6 | PELR_P_LORRI_ALICE_DEP_3 | Linear | ~65 | ~0.1 |
| 7 | PELR_PC_MULTI_DEP_LONG_2_01 | Linear | ~55 | ~0.1 |

We also note that all seven PCCs were seen close to the day-night terminator. The two PCCs identified in MVIC images (PCCs 3 and 4) are at locations within the encounter hemisphere, where high-resolution coverage is available for comparison at lower phase angles during approach. However, all five LORRI PCC candidates (PCCs 1, 2, 5, 6, 7) were located on the non-encounter hemisphere, which has much poorer low-phase spatial coverage. For all of the PCCs, their local time of day can be estimated based on their position on Pluto and the time the imagery was captured. The latitudes, longitudes, and local times are provided for each PCC in Table 3. Figure 5 shows their locations on a base map of Pluto, as well as the day-night terminator position near the time the images were acquired.

The seven PCCs spanned latitudes from 27.6 N to 19.0 S, with no clear preference for either the summer or winter hemisphere. We note that due to the high tilt of Pluto's rotational axis, the day-night terminator is located at only low- to mid-latitudes, and therefore we cannot rule out similar clouds at higher latitudes. However, cloud formation near the terminator would be consistent with the changing solar flux there that might promote cloud growth, as shown in our §2 results.

Table 3: Cloud Candidate Locations and Local Times on Pluto

| PCC | Image ID | UTC (DOY H:M:S) | Lat (deg) | E Long (deg) | Local Time (H:M) |
|---|---|---|---|---|---|
| 1 | PELR_P_MULTI_DEP_LONG_1 | 195 15:41:50 | +17.7 | 009.5 | 4:44 am |
| 2 | PELR_P_MULTI_DEP_LONG_1 | 195 15:41:50 | +27.6 | 359.5 | 4:03 am |
| 3 | PEMV_P_CharonLight | 195 12:11:24 | +2.5 | 205.0 | 5:12 pm |
| 4 | PEMV_P_CharonLight | 195 12:11:24 | -19.0 | 174.0 | 3:08 pm |
| 5 | PELR_P_LORRI_ALICE_DEP_3 | 196 00:03:20 | +05.6 | 005.8 | 5:47 am |
| 6 | PELR_P_LORRI_ALICE_DEP_3 | 196 00:03:20 | -09.8 | 021.2 | 6:49 am |
| 7 | PELR_PC_MULTI_DEP_LONG_2_01 | 195 19:04:45 | -07.8 | 033.5 | 6:51 am |

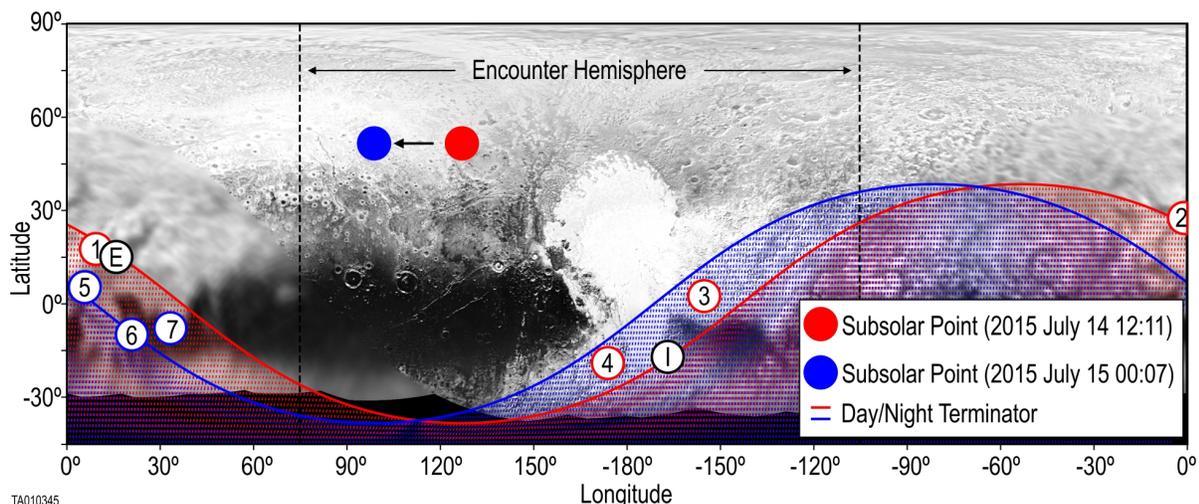

Figure 5. Cloud candidate locations on Pluto. The subsolar point and day-night terminator is indicated here at two times that bracket the acquisition times of the images containing the PCCs, at UTC 12:11 on 14 July 2015, corresponding to the acquisition time of the P_CharonLight image, and at UTC 00:07 on 15 July 2015, near the acquisition time of the P_LORRI_ALICE_DEP_3 image. All PCCs were detected in high phase post-closest approach imagery and were found near the limb of Pluto. As a result, all PCCs were found close to a terminator with either a local time near dusk (3, 4) or dawn (1, 2, 5, 6, 7). For reference, the New Horizons radio occultation Ingress (I) and Egress (E) locations are marked by black circles with these respective letter designations.

**3.3 Comparison with low-phase images**. PCCs 3 and 4 provide a unique opportunity for further analysis. This is because their locations were also imaged in lower phase, higher resolution LORRI and MVIC images. These images allow for (i) examination of the PCC region at different lighting conditions and spatial resolutions, and (ii) context to assess the effects of topography and surface albedo variation in the area.

Figure 6 shows the result of georeferencing for PCC 3, including both the original MVIC image and three panels of georeferenced information. While the cloud candidate is not seen in the low-phase imagery (Figure 6, bottom right), there does appear to be an association with topography suggesting either a possible surface explanation for the observed high phase brightening in that area (instead of a cloud), or a correlation between the formation of PCC 3 and the local topography. Figure 7 shows the result of georeferencing for PCC 4, including the original MVIC image and three panels of georeferenced information. This candidate is found in a region with scattered patches of bright surface albedo, and thus this PCC could be either a cloud or a surface feature; we cannot confirm/refute either possibility.

As Figures 6 and 7 reveal, georeferencing of PCCs 3 and 4 both suggests a correlation of each feature with a topographic depression that has a locally bright and smooth floor. Our analysis was inconclusive, however, as to whether these PCCs were simply caused by variability of the local surface scattering at high phase angles, or whether they are due to scattered light from low-lying clouds.

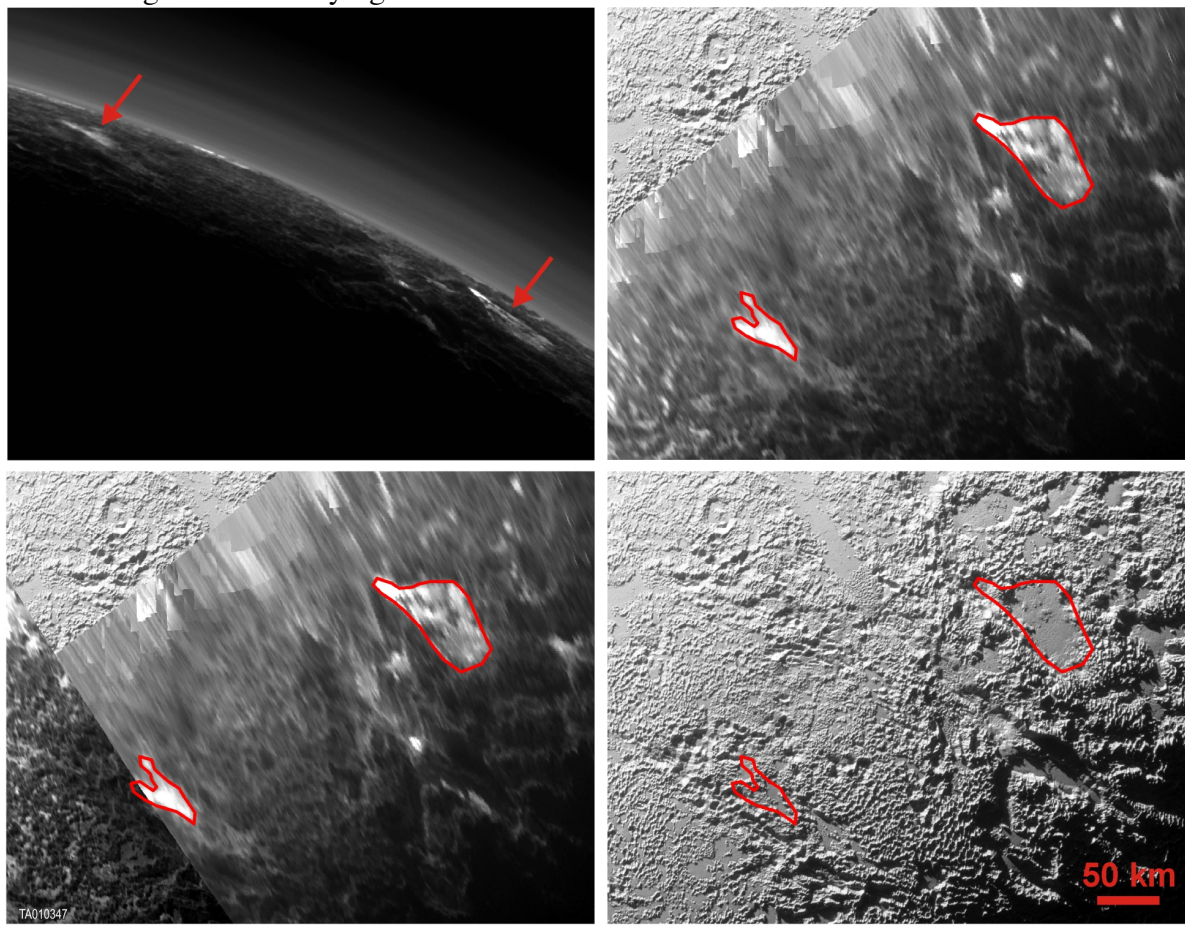

Figure 6. Georeferencing of PCC 3 to a simple cylindrical projection, centered at 210 E, 6 deg N. Top left: MVIC P_CharonLight image, cropped to show PCC 3 (left arrow) and a nearby bright patch (right arrow) on the surface of Pluto. Top right: MVIC P_CharonLight image over low phase basemap. Bottom, left: P_CharonLight overlaid with high phase MVIC image P_HiPhase_HiRes at the same location on Pluto. Bottom right: MVIC image P_MVIC_LORRI_CA, showing the same region at low-phase. In each panel the position of PCC 3 and the nearby bright patch are indicated by red arrows or outlines. The cloud candidate is observed in the high phase image but not the low phase image.

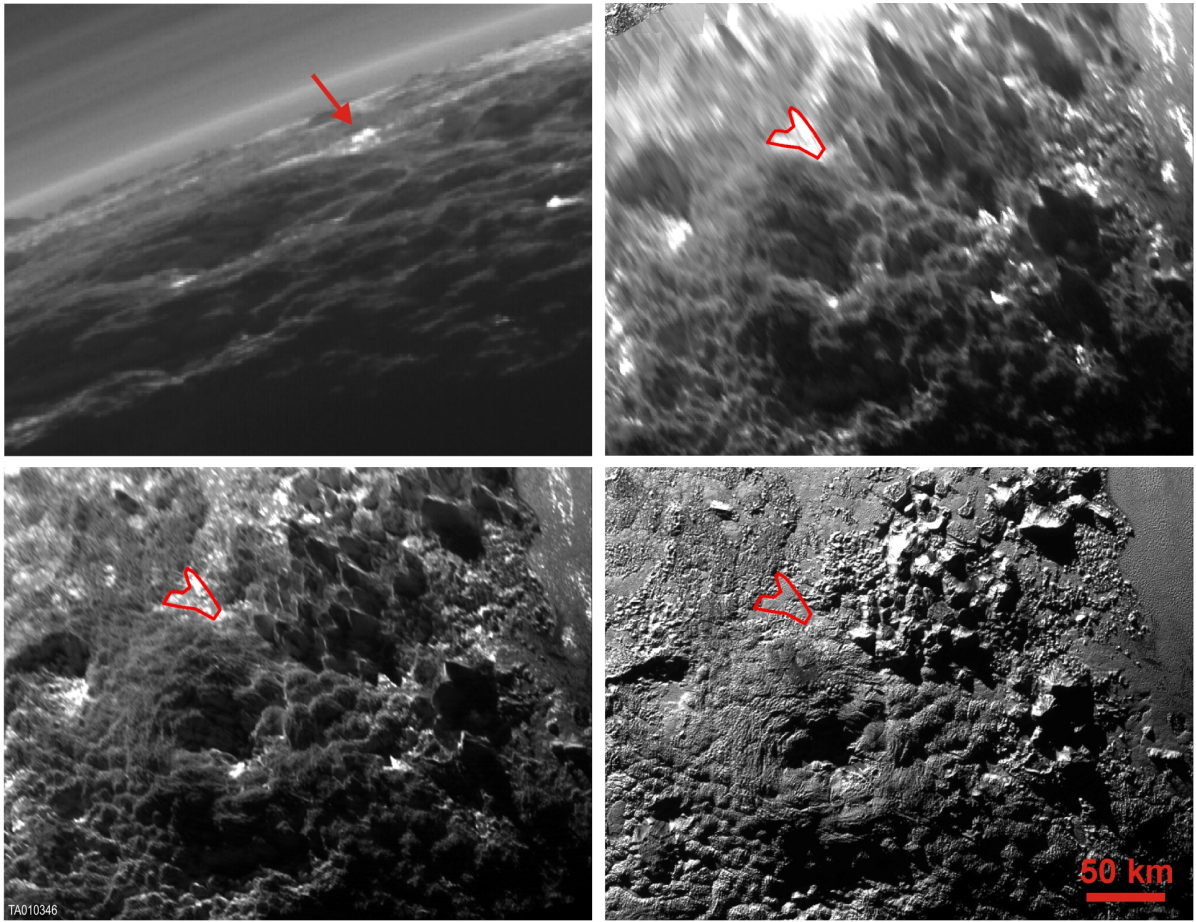

Figure 7. Georeferencing of PCC 4 to a simple cylindrical projection centered at Longitude 176 E, Latitude 18 deg S. Top, left: MVIC image P_CharonLight, cropped to show PCC 4. Top, right: georeferenced high-phase MVIC image P_CharonLight. Bottom, left: high phase MVIC image P_HiPhase_HiRes. Bottom, right: georeferenced MVIC image P_MVIC_LORRI_CA, showing the region at low-phase. In each case PCC 4's position is indicated by the red outline.

## 4. Cloud Candidate Attributes

The seven identified PCCs share several common attributes, which we point out and briefly discuss here. These include:

- **Small Size**. Both direct image analysis and georeferencing techniques confirm that all the PCCs are relatively small in size (10s of km maximum dimension).

- **Low Altitude**. All of our PCCs appear to lie near (within a few kilometers) of the surface. While their altitudes above the surface were not well constrained due to a lack of stereo imaging coverage at their locations, they additionally all appear to be low-lying relative to local topography, which would be also consistent with the predictions of the cloud condensation models discussed in §2.

- **Early/Late Local Time of Day**. All of the PCCs lie near local dawn or dusk. This is consistent with formation in cooler conditions where condensation is preferred, but could also be an observational selection effect resulting from the more favorable observing geometry near the planet's limb. A future orbiter mission should be able to distinguish between these possibilities.

- **Oblique Detection Geometry**. Finally, we note that all of the PCCs we identified were detected obliquely. Such detections may be the result of a selection effect in which enhanced path length may facilitate detection, indicating that additional PCCs may have escaped detection at more normal observing geometries.

## 5. Conclusions

We have used the New Horizons image dataset to search for discrete cloud features on Pluto. Only seven Pluto Cloud Candidates (PCCs) have been identified; all are small features. None can be absolutely confirmed to be a cloud, but all share attributes that are consistent with possible clouds.

Our more formal conclusions are as follows:

- At the current epoch, Pluto's atmosphere is almost entirely or entirely free of clouds.

- The seven PCCs we identified cannot be confirmed as discrete clouds because none are in regions where (i) stereo imaging, (ii) surface-lying shadows, or (iii) limb viewing geometry can be used to definitively demonstrate they are above ground features.

- Nonetheless, all seven PCCs share a range of common features (see §4) that circumstantially bolster the case that some or all may be low-lying clouds.
- No clouds were detected as discrete features above the limb, at backscattering (low phase angle) geometries, or away from Pluto's terminator. Because hazes were detected to altitudes as high as 220 km, we believe clouds could also have been be detected to such altitudes.

- ➢ Cloud condensation modeling we performed using vertical 1-D models indicates that our PCCs cannot be due to $N_2$ or CO condensation, but could be the result of condensation of any of several atmospheric trace species including HCN, $CH_4$, or any of several known light hydrocarbons.

Finally, we speculate that Pluto's seasonal and "super-seasonal" (Binzel et al. 2017; Earle et al. 2017; Stern et al. 2017) variations in cloud abundances, areal extents, and other attributes may be significant. Further modeling in this area is recommended. Specifically, a Pluto orbiter would be extremely useful to search for clouds more thoroughly in time and space than was possible during the brief reconnaissance flyby by New Horizons.

## 6. Acknowledgements

We thank NASA for financial support of the New Horizons project, and we thank the entire New Horizons mission team for making the results of the flyby possible. We also thank an anonymous referee for many helpful comments.